\documentclass[aps,prl,reprint,superscriptaddress, longbibliography]{revtex4-1}
\usepackage{H1}
\usepackage{amsmath}
\usepackage[pdftex,colorlinks=true]{hyperref}
\hypersetup{
citecolor = blue
}
\usepackage[normalem]{ulem}
\usepackage{amssymb}
\usepackage{amsthm}
\usepackage{amsfonts}
\usepackage{bbm}
\usepackage{array}

\begin{document}
\title{Signatures of integrability in the dynamics of Rydberg-blockaded chains}

\author{Vedika Khemani}
\affiliation{Department of Physics, Harvard University, Cambridge, MA 02138, USA}

\author{Chris R. Laumann}
\affiliation{Department of Physics, Boston University, Boston, MA 02215, USA}

\author{Anushya Chandran}
\affiliation{Department of Physics, Boston University, Boston, MA 02215, USA}

\graphicspath{{../FinalFigures/}}

\begin{abstract}
A recent experiment on a 51-atom Rydberg blockaded chain observed anomalously long-lived temporal oscillations of local observables after quenching from an antiferromagnetic initial state. 
This coherence is surprising as the initial state should have thermalized rapidly to infinite temperature.
In this article, we show that the experimental Hamiltonian exhibits non-thermal behavior across its entire many-body spectrum, with similar finite-size scaling properties as models proximate to integrable points. 
Moreover, we construct an explicit small local deformation of the Hamiltonian which enhances both the signatures of integrability \emph{and} the coherent oscillations observed after the quench.
Our results suggest that a parent proximate integrable point controls the early-to-intermediate time dynamics of the experimental system.
The distinctive quench dynamics in the parent model could signal an unconventional class of integrable system. 
\end{abstract}

\maketitle

\emph{Introduction---}
Remarkable experimental advances in the construction and control of synthetic quantum systems~\cite{BlochRMP,LukinNVReview, GirvinQED, Kinoshita:2006ve,Kondov:2011aa, Langen:2013aa,Schreiber:2015aa,Choi2016, Neill2016, Smith:2016aa, Kaufman:2016aa, Bernien:2017ab,LevPrethermal, GreinerMBLEntanglement2018} have revived interest in foundational questions about quantum thermalization and the emergence of statistical mechanics~\cite{PolkovnikovRMP,DAlessio:2016aa, Nandkishore14}.
Experiments have observed robust thermalization when the interactions are strong, localization and the concomitant persistence of initial state memory when spatial inhomogeneities are strong, and long-lived prethermal states near special integrable points~\cite{Kinoshita:2006ve,Kondov:2011aa,Langen:2013aa,Schreiber:2015aa,Choi2016, Neill2016, Smith:2016aa, Kaufman:2016aa, Bernien:2017ab,LevPrethermal, GreinerMBLEntanglement2018}.
 
Conventional wisdom holds that generic, strongly interacting isolated systems quickly reach local thermal equilibrium at infinite temperature, irrespective of the initial state~\cite{PolkovnikovRMP,DAlessio:2016aa, Nandkishore14}. 
It therefore came as a surprise when recent quench experiments in long Rydberg-blockaded atomic chains reported a strong initial state dependence in infinite temperature thermalization times~\cite{Bernien:2017ab}. 
In the Rydberg-blockaded regime, the effective dynamics occurs in a constrained manifold as adjacent atoms cannot simultaneously support Rydberg excitations.
The experiment observed long-lived coherent oscillations in local observables starting from a Ne\'el state with the maximum number of Rydberg excitations (called $|\mathbb{Z}_2\rangle$), but fast relaxation starting from a state with no Rydberg excitations ($|0\rangle$). 
What is the source of this coherence at infinite temperature?

In this Rapid Communication, we provide a partial answer by identifying signatures of integrability in the Hamiltonian controlling the time-evolution ($H_0$ in Eq.~\eqref{eq:H0}).
Specifically, we construct a deformation of $H_0$ that both magnifies the amplitude and lifetime of the coherent oscillations observed in quenches, \emph{and} monotonically enhances various spectral signatures of integrability. 
From our study, we hypothesize that a parent non-ergodic point that is proximate in parameter space to $H_0$ controls its short-time dynamics and small system-size eigenspectra. 
This parent point is a new model in constrained systems that has not been previously studied; although it is exhibits various signatures of integrability, the long-lived oscillatory response suggests that it could differ from more conventional examples of integrability in interesting ways.

The dynamics of constrained systems, and the possibility of unusual thermalization therein, has been studied in various contexts~\cite{OlmosLesanovsky, GarranFokker, GarrahanCleanMBL, Garrahanslow, Chandran:2016ab, Chen:2017aa}. Recent work~\cite{Turner:2018cr} has attributed the coherent oscillations in the Rydberg chain to presence of `quantum many-body scars', which are highly-excited many-body eigenstates that violate the eigenstate thermalization hypothesis~\cite{ShiraishiMori,BernevigAKLT,Turner:2018cr,LinMotrunichScars}, in loose analogy with the anomalous single-particle states that appear near certain classical periodic orbits in the semiclassical ($\hbar\rightarrow 0$) limit~\cite{HellerScars}. 
Our work helps firm up this analogy by identifying the parent integrable point controlling the dynamics with the ``classical model" and the magnitude of the deviation from this point with $\hbar$. 
Identifying and solving the parent model is therefore an intriguing route to analytically describing quantum scars and establishing their physical origin. 

\emph{Model---}
The Rydberg experiment~\cite{Bernien:2017ab} realizes a  quantum simulator composed of 51 qubits by coherently driving transitions between the ground state $|g\rangle \equiv |{\downarrow}\rangle$ and a highly excited Rydberg state $|r\rangle \equiv |{\uparrow}\rangle$ of neutral $^{87}$Rb atoms arranged in a linear array. 
Due to the strong van der Waals interaction between the excited atoms, it is energetically forbidden for neighboring atoms to be simultaneously excited, \emph{i.e.} states like $|\cdots {\uparrow}{\uparrow} \cdots \rangle$ are forbidden.
These Rydberg blockade constraints lead to a Hilbert space with dimension $\mathcal{D}$ given by the $(L+2)$'th Fibonacci number for a chain of length $L$.
Asymptotically, $\mathcal{D} \sim \left(\frac{1+\sqrt{5}}{2}\right)^L$. 
We assume that the constraint is strictly enforced so that the system never leaves the constrained manifold, and drop the smaller further neighbor interactions.
The effective Hamiltonian is then
\begin{equation}
H_0 = - \sum_{i=1}^L P_{i-1} X_{i} P_{i+1}, 
\label{eq:H0}
\end{equation}
where $X_i, Y_i, Z_i$ are Pauli spin 1/2 operators acting on the unconstrained Hilbert space and $P_i = (1- Z_i)/2$ is the projector onto the ground state of the atom at site $i$. 
With open boundary conditions, we define $P_0 = P_{L+1} = 1$. 
$H_0$ creates/destroys excitations at site $i$ only if the neighboring sites are down, preserving the constraints. 
Despite its apparent simplicity, the projectors make this model strongly interacting. 
Ref.~\onlinecite{Bernien:2017ab} reported two qualitatively different dynamical behaviors for the domain wall density ($\sum_i Z_i Z_{i+1}/L$) upon quenching to $H_0$ from the state with no Rydberg excitations, $|0\rangle = |{\downarrow}\rangle^{\otimes L}$, and the Ne\'el state, $|\mathbb{Z}_2\rangle = |{\downarrow}{\uparrow}{\downarrow}{\uparrow}\cdots\rangle$.  
Both states have energy $E=0$, corresponding to infinite temperature within the constrained Hilbert space.

The traceless Hamiltonian $H_0$ has time-reversal symmetry $\mathcal{T}$  and spatial inversion symmetry $\mathcal{I}$  about the central site/bond.
In addition, $H_0$ anticommutes with the operator $\mathcal{P} = \prod_{i=1}^L Z_i$ so that the eigenenergies come in $\pm E $ pairs for $E \neq 0$. 
The zero energy manifold is highly degenerate ($\dim \ker H \sim \sqrt{\mathcal{D}}$) due to these symmetries  \cite{Turner:2018cr, Schecter}.
This degeneracy is not entropically relevant and does not play an important role in what follows.

The complete set of symmetry-preserving local deformations of $H_0$ up to range four is captured by the deformed Hamiltonian,
\begin{align}
H&= H_0 - \sum_i h_{XZ} (P_{i-1} X_i P_{i+1} Z_{i+2} + Z_{i-2} P_{i-1} X_i P_{i+1}) \nonumber \\
& - h_{YZ} (P_{i-1} Y_i P_{i+1} Z_{i+2} + Z_{i-2} P_{i-1} Y_i P_{i+1}). \label{Eq:HamDeformed}
\end{align}
We have included a time-reversal breaking $h_{YZ}$ term as well for future comparison.
We will see that the proximate near integrable point lies in this expanded parameter space. 
A more detailed derivation of the Hamiltonian in \eqref{Eq:HamDeformed} is given in Appendix A. 

Other deformations of $H_0$ have been studied before~\cite{Fendley:2004aa, Lesanovsky2012,Lesanovsky:2012aa, SachdevGirvinMott}; notably, these include diagonal terms such as $\sum_i Z_i$ which break $\mathcal{P}$. 
Although there are known integrable lines in these models~\cite{Fendley:2004aa} (including the famous Golden Chain~\cite{Feguin2007}), we do not numerically find that they are relevant to explaining the quench dynamics governed by $H_0$. 
A possible reason is that these deformations impart different energy densities to the $|0\rangle$ and $|\mathbb{Z}_2\rangle$ states, moving them away from infinite temperature and from one another.

\begin{figure}
\includegraphics[width=\columnwidth]{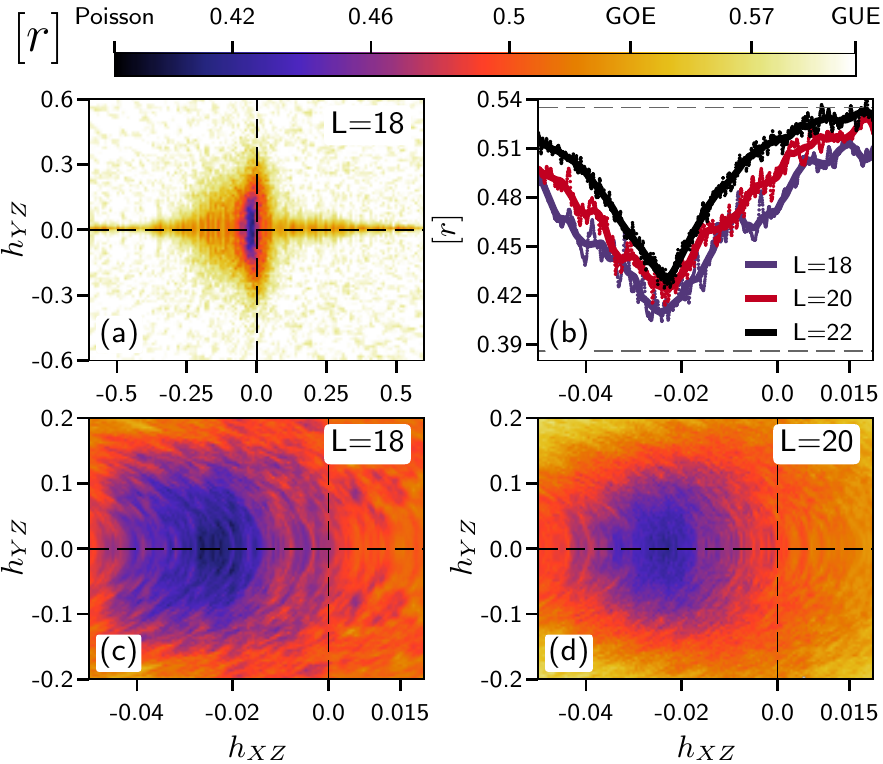}
\caption{
Color plots of the mean level spacing ratio $[r]$ vs $h_{XZ}$ and $h_{YZ}$ for the model \eqref{Eq:HamDeformed}. 
(a) shows robust thermalization to the appropriate GOE/GUE ensemble in most of the parameter space.  
There is an integrable looking region in the vicinity of the origin $H_0$ near $h_{XZ} \approx -0.02$; this is clearer in the zoomed-in panels (c) and (d). 
(b) shows $[r]$ as a function of $h_{XZ}$ and $L$ for $h_{YZ}=0$, showing both a dip in $[r]$ towards the integrable value as a function of $h_{XZ}$, and a gradual drift of the dip value towards thermal with increasing $L$ --- indicating that the exact integrable point requires further deformations. }
\label{Fig1:Meanr}
\end{figure}

\emph{Level Statistics---}
To start, we explore the infinite-temperature dynamical properties of the deformed model \eqref{Eq:HamDeformed} using the level spacing ratio $r_n$, defined as $r_n \equiv \textrm{min}(\Delta E_{n+1}/\Delta E_{n}, \Delta E_n/\Delta E_{n+1})$ where $\Delta E_n = E_n - E_{n-1}$ and $E_n$ is the $n^{th}$ energy eigenvalue \cite{Oganesyan:2007aa,Atas:2013aa}.  
In thermal systems, spectrally averaged $r_n$ ($[r]$) flows with system size to the Gaussian Unitary Ensemble (GUE) value $0.6$ when time reversal is broken ($h_{YZ}\neq0$) and the Gaussian Orthogonal Ensemble (GOE) value $0.53$ otherwise\cite{Atas:2013aa}.
In an integrable system with extensively many conservation laws, energy levels do not repel and $[r] \rightarrow 0.39$ corresponding to Poisson statistics. 
\emph{Near} an integrable point, $[r]$ may look intermediate between the Poisson and thermal values, but flows towards the thermal value with increasing system size\cite{MillisIntegrabilityCrossover}. 
Prior work on $H_0$ has observed a very slow flow of $[r]$ towards the GOE value with increasing $L$~\cite{Turner:2018cr, Turner_pc}. 

Fig.~\ref{Fig1:Meanr}(a) shows $[r]$ averaged over the middle third of the spectrum and the $\mathcal{I}=\pm1$ inversion sectors as a function of $h_{XZ}$ and $h_{YZ}$. 
Except for a small region in the vicinity of ---but not-centered on--- $H_0$, $[r]$ comes close to its random matrix value, suggesting robust thermalization. 
This confirms that the presence of constraints and the zero-energy degeneracy do not impede thermalization \cite{Chandran:2016ab, Chen:2017aa}.  
Panels (c) and (d) zoom into the region near $H_0$, revealing strong signatures of integrability ($[r] \approx 0.39$). 
The apparently integrable region shrinks toward $h_{XZ} \approx -0.02, h_{YZ} \approx 0$ with increasing $L$, suggesting that there is a near integrable \emph{point} (rather than an integrable manifold), which controls the scaling of $[r]$ at $H_0$ for accessible system sizes.
The dramatic decrease in $[r]$ towards the Poisson value takes place over a very small $\Delta h_{XZ} \approx 0.02$; this sensitivity is symptomatic of proximity to integrability. 

\begin{figure*}[t]
\includegraphics[width=\textwidth]{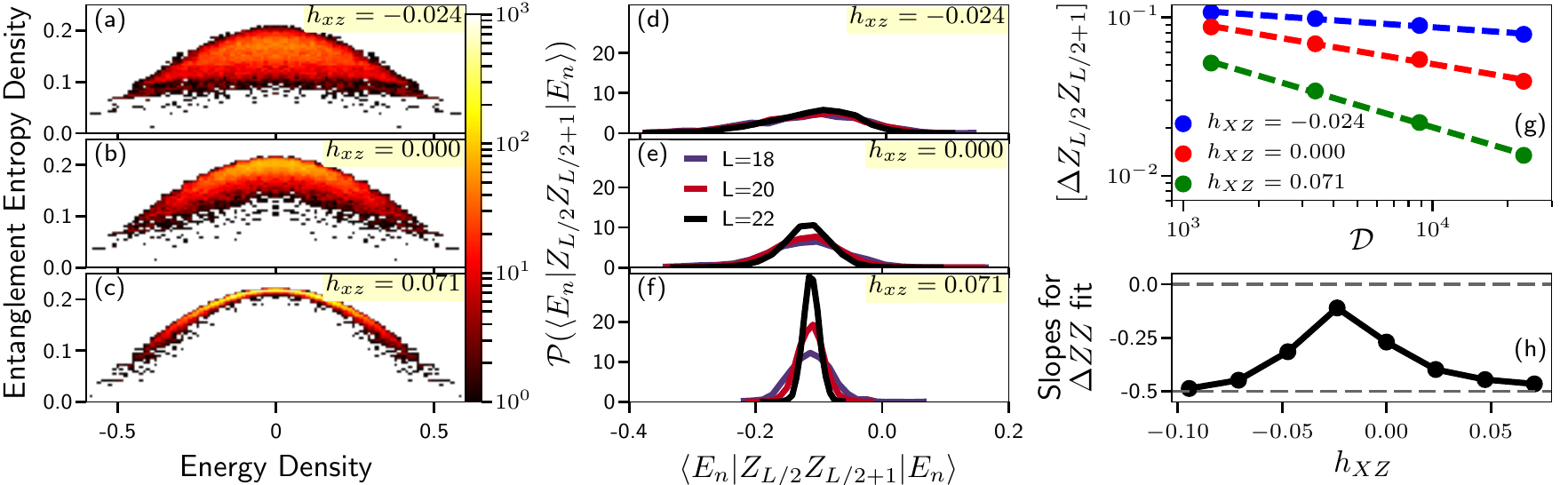}
\caption{(a) - (c): Half-chain entanglement entropy (EE) density of eigenstates plotted against energy density for $L=22$ and $h_{YZ}=0$ (a-c).   (d)-(f) show the distribution of EEVs across the middle third of the eigenspectrum for the domain-wall operator at the center of the chain, showing strong narrowing with increasing $L$ for $h_{XZ}\simeq 0.07$ (f) but almost no narrowing for $h_{XZ}\simeq -0.02$ (d).   (g) shows the scaling of fluctuations of EEVs between neighboring eigenstates \eqref{eq:ETHDelta} as a function of Hilbert space dimension $\mathcal{D}$. This quantity narrows slower than the ETH prediction $\mathcal{D}^{-1/2}$ for $H_0$ and $H_{\rm int}$.  (h) shows the scaling of the fluctuations $\Delta$ with $\mathcal{D}$ as a function of $h_{XZ}$. The narrowing is slowest for $h_{XZ} \simeq -0.02$ but increases  towards the ETH value of $-1/2$ on perturbing $h_{XZ}$ away from this value. The intermediate values of the slope (approximately $-1/4$ for $H_0$) are consistent with proximity to integrability\cite{HaqueETH}.  }
\label{Fig:ETHStudy}
\end{figure*}

As the level-statistics data is relatively insensitive to the breaking of time-reversal by $h_{YZ}$, we set $h_{YZ}=0$ henceforth.
Fig.~\ref{Fig1:Meanr}(b) shows $[r]$ as a function of $h_{XZ}$ for different system sizes $L$. 
While there is a pronounced dip in $[r]$ towards the Poisson value near $h_{XZ} \approx -0.02$, the value of $[r]$ near this dip shows a slow drift towards the GOE value with increasing $L$. 
This implies that while our chosen deformation makes $H_0$ look more integrable, the exact proximate integrable point likely does not strictly live in the two-dimensional parameter space explored by \eqref{Eq:HamDeformed}. 
Given the smallness of the optimal $h_{XZ}$, it is possible that we have only found the leading terms of a quasi-local integrable Hamiltonian which includes a hierarchy of additional longer range terms with smaller amplitudes. 

Below, we use the deformed model with $h_{XZ}=-0.0236$ (where $[r] \approx 0.39$) as a proxy for the parent model of $H_0$ at the numerically accessible system sizes, and the model with $h_{XZ} = 0.0708$ (where $[r] \approx 0.53$) as an example of a strongly thermalizing point. We denote the Hamiltonians at these points as $H_{\rm int}$ and $H_{\rm th}$ respectively.

\emph{Eigenstate entanglement entropy---}
The eigenstate thermalization hypothesis (ETH) states that thermalization occurs at the level of individual eigenstates~\cite{Jensen:1985aa,Deutsch, Srednicki, Rigol}.
When the ETH holds, systems locally thermalize irrespective of the initial state.
A convenient observable-independent diagnostic of the ETH is the half-chain von Neumann entanglement entropy (EE) evaluated in eigenstates (Fig.~\ref{Fig:ETHStudy} (a-c)).
The ETH implies that this quantity coincides with the thermal entropy density $S(E)$, and is accordingly a smooth function of energy density as $L \to \infty$. 
This is clearly seen in the narrow scatter in Fig.~\ref{Fig:ETHStudy}(c) for the EE evaluated in $H_{\rm th}$.

In contrast, distribution of observables can be extremely broad in integrable systems, with a width that does not narrow even as $L\rightarrow\infty$~\cite{DAlessio:2016aa}.  
Eigenstates in integrable systems are labeled by extensively many conserved quantities whose sectors coexist at the same energy density. 
The number of states in a given sector ranges from $O(1)$ to $O(L)$ to $\exp(O(L))$, and the corresponding EE can range from $0$ to $O(\log(L))$ to $O(L)$ all at the same energy.
The broad scatter of the EE evaluated in $H_{\rm int}$ (Fig.~\ref{Fig:ETHStudy}(a)) is characteristic of such integrable systems. 

The EE distribution in Fig.~\ref{Fig:ETHStudy}(b) for $H_0$ clearly lies between that of $H_0$ and $H_{\rm th}$. It is narrower than that of $H_{\rm int}$, but exhibits outlier states with small EE near infinite temperature even at $L=22$.
We hypothesize that these outlier states (dubbed many-body scars in Refs.~\cite{Turner:2018cr, TurnerScars2}) at $H_0$ are thus a finite-size shadow of the low-entropy conserved sectors in the proximate integrable parent model. 
Finally, notice that this dramatic change in eigenstate properties across (a)-(c) takes place over a very small range of $\Delta h_{XZ}$, consistent with the parameter sensitivity seen near integrability.  

\emph{Finite-size ETH scaling---}
To establish proximity to integrability at $H_0$, we turn to a quantitative finite-size ETH scaling study.  
The ETH hypothesizes that few-body observables $\hat{O}$ in eigenstates depend smoothly on energy $E$ with small fluctuations\cite{Srednicki}. More precisely,
\begin{align}
\Delta O_{n} \equiv |\langle E_n | \hat{O} | E_n \rangle - \langle E_{n-1} | \hat{O} | E_{n-1} \rangle| \sim \frac{1}{\sqrt{e^{S(E_n)}}}.
\label{eq:ETHDelta}
\end{align}
At infinite temperature, $S(E=0) = \log \mathcal{D}$ so, $\Delta O_{n} \sim 1/\sqrt{\mathcal{D}}$ \cite{Kim:2014aa}.
In contrast, neighboring energy eigenstates in integrable systems can belong to different conservation law sectors and thus have very different expectation values, so that $\Delta O_{n} \sim \mathcal{D}^0$. 

Fig.~\ref{Fig:ETHStudy} (d-f) shows the distributions of the eigenstate expectation values (EEVs) for $O =Z_{L/2} Z_{L/2+1} $ across eigenstates in the middle third of the spectrum.
The width of this distribution shows no/weak/strong narrowing with increasing $L$ for $H_{\rm int}/H_0/H_{\rm th}$ respectively.
Fig.~\ref{Fig:ETHStudy}(g) quantifies the scaling of the width with $\mathcal{D}$ using the spectrally averaged $[\Delta O_n]$.
Only $H_{\rm th}$ exhibits the $\mathcal{D}^{-1/2}$ scaling predicted by Eq.~\eqref{eq:ETHDelta}; the other two Hamiltonians exhibit significantly slower scaling with $\mathcal{D}$.

Fig.~\ref{Fig:ETHStudy}(h) plots the slopes $\alpha$ of the curves in panel (g) $[\Delta O] \sim \mathcal{D}^\alpha$ as a function of $h_{XZ}$.
The slopes decreases \emph{monotonically} from close to zero at $h_{XZ} = -0.02$, reaching the ETH value of $-0.5$ at large $|h_{XZ}|$.
This behavior is a finite-size effect and has been numerically observed near known integrable points \cite{HaqueETH}.
Perturbing away from the integrable point where $\alpha = 0$ by $\epsilon$ generates a scattering length $\ell(\epsilon)$ beyond which the states in different conserved sectors mix. 
That is, $\alpha$ smoothly crosses over from $0$ to $-0.5$ at the value $\epsilon_c$ such that $\ell(\epsilon_c) = L$.
As $L \to \infty$, $\epsilon_c \to 0$ and $\alpha = -0.5$  for $\epsilon \neq 0$
Fig.~\ref{Fig:ETHStudy}(h) thus provides strong evidence that $H_0$ exhibits anomalous EEV scaling at the numerically accessible system sizes due to its proximity to a parent integrable point.

\begin{figure}
\includegraphics[width=\columnwidth]{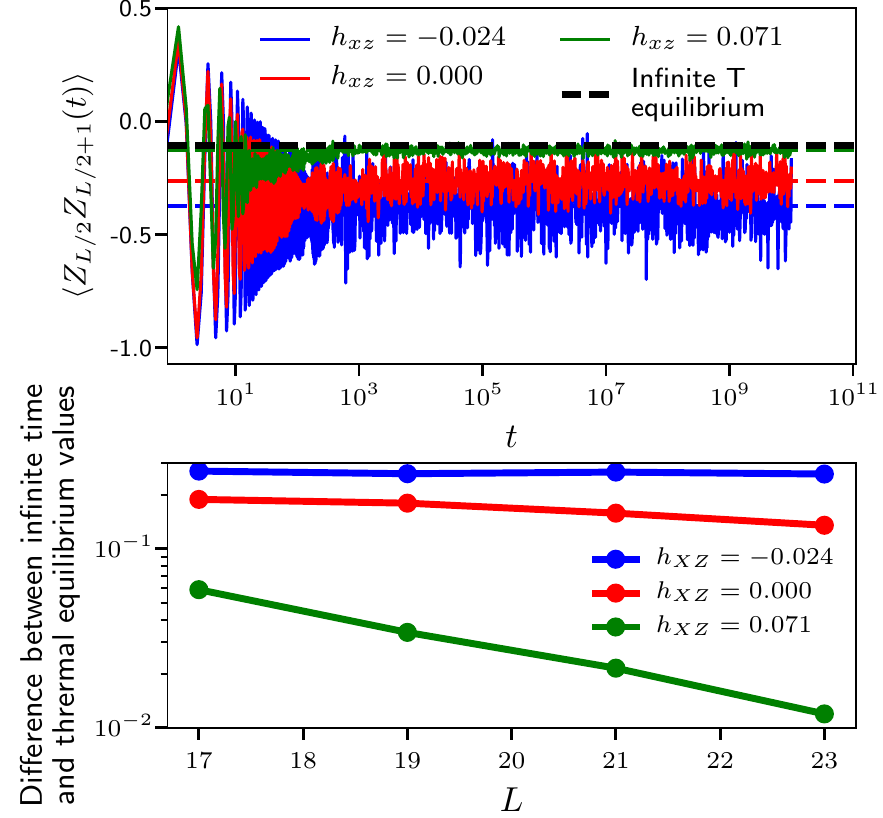}
\caption{(a): Time dynamics for a local domain wall operator starting from the $|\mathbb{Z}_2\rangle$ state for $L=22$ and three different quench Hamiltonians $H_{\rm int}$, $H_0$ and $H_{\rm th}$. The dynamics shows large initial oscillations in all cases, with the amplitude at $H_{\rm int}$ being much larger than that at $H_0$ or $H_{\rm th}$.  (b) shows the difference between the late time diagonal ensemble value and the canonical equilibrium prediction as a function of $L$ for the three different quench Hamiltonians. While this difference decreases slowly with $L$ for $H_0$, we see no significant flow for $H_{\rm int}$ and a fast decrease for $H_{\rm th}$. }
\label{Fig:TimeDynamics}
\end{figure}

\emph{Dynamics---} 
The previous numerical results show that the bulk spectral and eigenstate behavior of $H_0$ are controlled by a proximate model with strong signatures of integrability.
Fig.~\ref{Fig:TimeDynamics}(a) shows that the experimentally intriguing coherent oscillation after quenching from $|\mathbb{Z}_2\rangle$ is also strongly correlated with the proximity to $H_{\rm int}$. 
Under evolution by each of the three representative Hamiltonians, $H_{\rm int}, H_0, H_{\rm th}$, the domain wall density relaxes to a stationary value after a long-lived oscillation.
The amplitude and duration of the oscillation significantly increases from $H_{\rm th}$ to $H_0$ and then to $H_{\rm int}$, despite only a modest change in $h_{XZ}$.
Likewise, the difference between the late time value (dashed) and the expected thermal value (black dashed) increases, as do the fluctuations about that value.
In contrast, all three models quickly relax to thermal equilibrium upon quenching from the state $|0\rangle$ (not shown).

The proximity to integrability explains both the \emph{late} time dynamics and sensitivity to initial conditions.
In the absence of accidental spectral degeneracies, a local observable $\langle O \rangle$ relaxes to a late-time value given by the \emph{diagonal ensemble} $O_d = \sum_n |c_n|^2 \langle E_n | O | E_n \rangle$ after quenching from $|\psi_0\rangle = \sum_{\alpha} c_n |E_n\rangle$ due to dephasing~\cite{DAlessio:2016aa}.
In thermalizing systems, $O_d$ agrees with its thermal Gibbs ensemble  value $O_{\rm GE}$ at a temperature set by the energy of $|\psi_0\rangle$ as $L\to\infty$. 
In integrable systems, $O_d$ instead agrees with a Generalized Gibbs ensemble (GGE), which is parameterized by an additional  $O(L)$ \emph{initial state dependent} chemical potentials for each of the additional conservation laws~\cite{Jaynes:1957aa, Jaynes:1957ab, Rigol:2007cr}.
Thus, in the $L\to\infty$ limit, $|O_d - O_{\rm GE}|$ can remain non-zero in an integrable system for initial states with non-zero chemical potentials. 

Fig.~\ref{Fig:TimeDynamics}(b) shows the diagonal ensemble value of the domain wall density converging rapidly with $L$ to the thermal Gibbs value for $H_{\rm th}$, but not at all for $H_{\rm int}$ at accessible sizes.
This is consistent with $H_{\rm th}$ being thermalizing and $H_{\rm int}$ being very close to integrable. 
We note that the trend in the scale of fluctuations around the late time value seen in Fig.~\ref{Fig:TimeDynamics}(a) is also consistent with these ensembles (finite-size scaling analysis not shown).

The difference between $O_d$ and $O_{\rm GE}$ converges slowly with $L$ for $H_0$, which can be understood by  proximity to integrability.
For $L < \ell(\epsilon)$, the conservation laws at the parent model approximately hold even for infinite time so that $O_d$ differs from $O_{\rm GE}$. 
For $L > \ell(\epsilon)$, this difference crosses over toward zero as the different sectors starting mixing, as visible in Fig.~\ref{Fig:TimeDynamics}(b) for the largest sizes. 
We note that we expect a similar decay for $H_{\rm int}$ at larger $L$ than shown, as it is not exactly integrable either.
Finally, even as $L \to \infty$, the \emph{finite-time} dynamics up to a crossover time scale $\tau(\epsilon)$ are still governed by the parent model, as is well known in the study of prethermal phenomena~\cite{DAlessio:2016aa,PrethermalMartin,PrethermalEssler}
\footnote{We note that the coherence times observed in \cite{Bernien:2017ab} are experimentally limited by the further neighbor van der Waals interactions, which is a much larger deformation of the parent model than $H_0$ with a correspondingly shorter $\tau(\epsilon)$.}.

A detailed description of the \emph{early time} oscillatory behavior in $H_0$ is still an outstanding challenge. 
Close to integrable systems, which have quasiparticle descriptions even at infinite temperature, long-lived quench oscillations can arise due to heavy, slowly dispersing quasiparticles~\cite{Delfino_2014}. 
More exotic integrable models could even exhibit exactly periodic modes. 
Understanding this in analytic detail, however, will require a more complete solution of the parent model.   
This would also permit analytic control of the time and length scales generated by the deformation back to $H_0$. In this vein, we note that a recent preprint~\cite{ChoiPerfectScars} found that the amplitude of early-time oscillations in the deformed model introduced by us in \eqref{Eq:HamDeformed} is maximized at $h_{XZ} \simeq -0.04, h_{YZ} \simeq 0$. Although this point does not coincide with the integrable point at $h_{XZ} \simeq -0.02$, Fig.~\ref{Fig:ETHStudy} shows that the entire many-body  eigenspectrum at  $h_{XZ} \simeq -0.04$ is still strongly influenced by the integrable point at the numerically accessible system sizes. An intriguing possibility is that the weak breaking of integrability preserves conservation laws in a small subspace of the full Hilbert space, and thus gives rise to exact scar states via the mechanism outlined in Ref.~\cite{ShiraishiMori}.
We defer an exploration of this connection to further work.

\emph{Discussion---} 
We have presented evidence that a parent non-ergodic model with strong signatures of integrability controls the properties of the Rydberg-blockaded chain Hamiltonian $H_0$.
The \emph{entire} many-body spectrum of $H_0$ violates finite-size ETH scaling in a manner consistent with proximity to integrability. Strikingly, the coherent post-quench oscillations observed experimentally~\cite{Bernien:2017ab} are enhanced by deformation toward the parent model. 
A consistent explanation for the ``scar states"  observed in Ref.~\cite{Turner:2018cr} is that these are the finite-size shadow of low entropy conserved sectors of the parent model. 
Unless we are exactly at the integrable point, we expect these special states to disappear with increasing system size --- even though finite time properties continue to be governed by the parent Hamiltonian. 

Our work raises a number of interesting questions. 
First, what is the exact parent Hamiltonian? 
Likely, we have only found the first terms in a quasilocal expansion of a previously unknown exactly integrable Hamiltonian with direct experimental implications. Finding this exact integrable-looking point and understanding its properties is an important direction for future study.
Second, is the integrability ``conventional'' for one dimensional chains? 
Intriguingly, the sign of $h_{XZ}$ is consistent with having a classical two-dimensional statistical description without a sign problem~\footnote{This can be seen by rewriting the deformation term as $P_{i-1}X_i P_{i+1} P_{i+2} + ({\rm inverted})$ with a small renormalization of $H_0$.}. 
On the other hand, state-dependent, long-lived quench oscillations have not been reported in known integrable models.
Understanding their analytic origin will reveal either an unconventional class of integrability, or new dynamical regimes within existing models.

\emph{Acknowledgements---} We thank F. Burnell, S. Choi, W.W. Ho, D. Huse, P. Fendley,  M. Lukin, R. Moessner, H. Pichler,  S. Sondhi and A. Vishwanath for many enlightening discussions, and C.~J.~Turner and Z.~Papic for sharing unpublished spectral statistics data for $H_0$. 
VK is supported by the Harvard Society of Fellows and the William F. Milton Fund. 
CRL acknowledges support from the Sloan Foundation through a Sloan Research Fellowship and the NSF through grant No. PHY-1656234. 
AC acknowledges support from the NSF through grant No. DMR-1752759.
Any opinion, findings, and conclusions or recommendations expressed in this material are those of the authors and do not necessarily reflect the views of the NSF. 

\emph{ Note Added---} Recently, we became aware of two  complementary works~\cite{TurnerScars2,TDVPScars} that study the parent model $H_0$, particularly in relation to its connection with quantum scars; the first uses a forward scattering approach to construct low-entanglement eigenstates of $H_0$~\cite{TurnerScars2}, while the latter uses a matrix product state approach to derive closed periodic orbits~\cite{TDVPScars}.

\appendix

\section{Appendix A: Deformations up to Range 4}
\label{sec:AppModel}
The number of independent operators acting on the constrained space grows asymptotically as $\phi^{2L}$, where $\phi = (1+\sqrt{5})/2$ is the Golden ratio.
Up to range four, there are 11 independent operators:
(1)~$\sum_i Z_i$, 
(2)~$\sum_i Z_i Z_{i+2}$, 
(3)~$\sum_i Z_i Z_{i+3}$, 
(4)~$\sum_i P_{i-1} X_i P_{i+1}$, 
(5)~$\sum_i P_{i-1} Y_i P_{i+1}$, 
(6)~$\sum_i P_{i-1} X_i P_{i+1}Z_{i+2}$, 
(7)~$\sum_i Z_{i-2} P_{i-1} X_i P_{i+1}$, 
(8)~$\sum_i P_{i-1} Y_i P_{i+1}Z_{i+2}$, 
(9)~$\sum_i Z_{i-2} P_{i-1} Y_i P_{i+1}$, 
(10)~$\sum_i P_{i-1} S^+_i S^-_{i+1}P_{i+2}$, 
(11)~$\sum_i P_{i-1} S^-_i S^+_{i+1}P_{i+2}$. 
Deformations of $H_0$ that are diagonal in the $z$-basis such as (1) and (2) have been studied before in Ref.~\cite{Fendley:2004aa}.
`Hopping' deformations (10) and (11) have also been previously studied \cite{Cheong:2009aa}. 
All of these deformations contain integrable manifolds in parameter space. 
However, they do not anticommute with $\mathcal{P}$ and we have numerically observed that breaking $\mathcal{P}$ rapidly leads to thermalization in models perturbatively adjacent to $H_0$.

The minimal deformations of $H_0$ at this range which respect all of the symmetries described in the main text are captured by:
\begin{align}
H_1&= H_0 + \sum_i h_{XZ} (P_{i-1} X_i P_{i+1} Z_{i+2} + Z_{i-2} P_{i-1} X_i P_{i+1})
\end{align}
In the main text, we have also included terms (8) and (9) which break time reversal in order to illustrate the crossover from GOE to GUE in the level statistics for comparison.
We note that term (5) can be absorbed into $H_0$ by a rotation about the $Z$-axis.

\bibliography{global,paper-master}

\end{document}